\documentclass[12pt]{article}

\usepackage{graphicx}

\begin{document}

\title{\Large \bf Recent Results from the JLab Spin Physics Program }
\author{\large K. Slifer \bigskip \\
{\it University of Virginia} 
}

\maketitle


\begin{center}
{\bf Abstract}\\
We present here select recent results from the Thomas Jefferson National Laboratory  Spin Physics program,
along with the perspective on some upcoming experiments.
\end{center}

\section{EG4:
The Proton/Deuteron GDH Sum at Low $Q^{2}$}

\begin{figure}\begin{center}\includegraphics[width=0.9\textwidth]{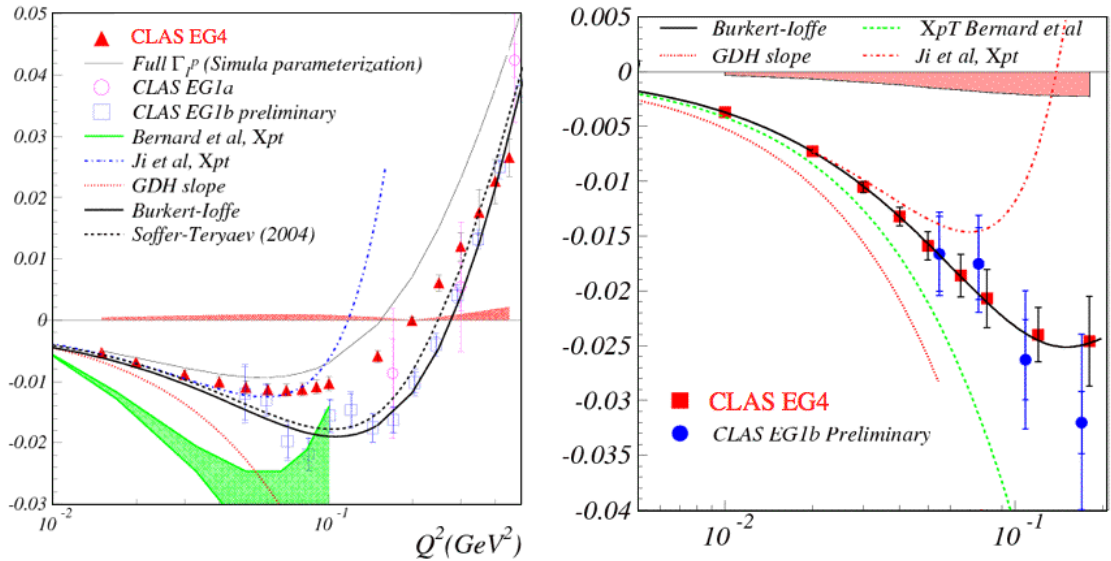}
\caption{\label{PREDD} {\bf Left:} Projected uncertainty of proton $\Gamma_1$.
{\bf Right:} Projected uncertainty of deuteron  $\Gamma_1$.
The band represents the systematic uncertainty, while the error bars on the
   points are statistical only.
   The curves from
   Bernard {\it et al.} and
   Ji {\it et al.} are $\chi$PT calculations.
   The curve from Burkert-Ioffe
   and Soffer-Teryaev are phenomenological models. The preliminary EG1B
   data are also shown for comparison.
                                                                 }
\end{center}\end{figure}

The Gerasimov-Drell-Hearn(GDH) sum
rule~\cite{gdh}
for real photon scattering
from a target of arbitrary spin $\mathcal{S}$ reads:
\begin{eqnarray}
\int_{\nu_{th}}^\infty \frac{ \sigma_{P}(\nu) - \sigma_{A}(\nu)}{ \nu}d\nu
&=& -4\pi^2\alpha~ \mathcal{S} \left(\frac{\kappa}{m}\right)^2
\label{GDHSR}
\end{eqnarray}
where  $\sigma_{P}$ and $\sigma_{A}$ represent the cross
section for photoabsorption with the photon helicity parallel or
anti-parallel to the target spin in its maximal state. The target mass and
anomalous  magnetic moment are represented by
$m$ and $\kappa$ respectively, and  $\nu$ is the photon energy.
Extending the sum rule
to finite $Q^2$ provides a powerful tool to investigate the spin structure of the nucleon.
For a hadronic target the
generalization~\cite{ji&osborn}
can be written:
\begin{eqnarray}
 \label{JISUMRULE}
 \overline{S_1}(0,Q^2)  =     \frac{8}{Q^2} \int_0^{1-\epsilon} g_1(x,Q^2) ~dx
                       \equiv \frac{8}{Q^2} \overline{\Gamma}_1
\end{eqnarray}
Here $g_{1}$ is the spin structure function,
$S_1$ is the forward Compton amplitude, and
the overbar represents exclusion of the elastic contribution.
The forward Compton amplitudes can be evaluated
in Chiral
Perturbation theory at low Q$^{2}$, or via the higher twist expansion at large
Q$^{2}$. Lattice QCD holds the promise of eventually evaluating these moments
at any Q$^{2}$.

EG4~\cite{EG4}  ran successfully in JLab's Hall B in early 2006.
The goal was to extend our knowledge of the proton and deuteron(neutron) spin structure
to the lowest possible JLab momentum transfer
($0.015<Q^{2}<0.2$ GeV$^{2}$) in the resonance region ($W<2$ GeV).

In order to extract $g_1$ we performed an absolute (polarized) cross
section measurement.  For this purpose, a new \v{C}erenkov counter
was built and commissioned by the INFN Genova  group.
This new detector was designed for
high electron detection efficiency (of the order 99.9\%)
and a high pion rejection
ratio (of the order 10$^{-3}$).
We also utilized the JLab-UVA polarized target.
This target exploits the
Dynamical Nuclear Polarization (DNP) technique to polarize a solid ammonia insert
maintained in a liquid helium bath at 1 K in a 5 Tesla
field.  

In Fig.~\ref{PREDD}, we show the expected precision for the proton and deuteron extended
GDH sum, along with several theoretical predictions.
The high precision data will also allow an
extraction of the neutron extended GDH sum.  In addition to testing
$\chi$PT and lattice calculations, we will indirectly test the
the real photon GDH sum rule, by extrapolating to $Q^2=0$.

\section{$\delta_{LT}$ : The Generalized LT Spin Polarizability}

In recent years, Chiral Perturbation Theory ($\chi$PT) has emerged
as {the} effective theory of QCD at low energy.
It has also been used as a powerful tool to help Lattice QCD
(LQCD) to extrapolate to the physical region.
In light of this, it is critical to have benchmark tests of the reliability of $\chi$PT
calculations.
The JLab experimental program
on the spin structure of the nucleon
has extracted the spin structure functions $g_1^n$, $g_2^n$ and $g_1^p$ and their
moments over a wide kinematic range.
These moments have proven to be powerful tools
to test QCD sum rules and $\chi$PT calculations.
However, at the low $Q^2$ relevant to $\chi$PT, data on the
$g_2^p$ structure function is conspicuously absent.
Currently, the lowest momentum transfer that has been investigated is
$Q^2\approx 1.3$ GeV$^2$ by the RSS collaboration~\cite{PRL}.

The absence of transverse
data is particularly unsatisfying given the intriguing results found in the
transverse neutron data: The SLAC E155
collaboration 
found a three sigma violation of the proton Burkhardt-Cottingham (BC) sum rule at
$Q^2=5.0$ GeV$^2$, while the JLab E94010
collaboration 
found that the
neutron BC sum rule held below
$Q^2=1.0$ GeV$^2$.
Even more compelling, it was found that state-of-the-art NLO
$\chi$PT calculations are in agreement
with the neutron data for the generalized polarizability $\gamma_0^n$ at $Q^2=0.1$ GeV$^2$, but
exhibit a significant discrepancy~\cite{Amarian:2004yf} with the longitudinal-transverse polarizability
$\delta_{LT}^n$.
This is particularly surprising since
$\delta_{LT}$ is
insensitive to the $\Delta$ resonance contribution
which is not well under control in the $\chi$PT calculations.
For this reason, it was believed that $\delta_{LT}$ should be more suitable than
$\gamma_0$ to serve as a testing ground for the chiral dynamics of
QCD~\cite{ber,van02}.
It is natural to ask if this discrepancy exists in the proton case,
and determining
the isospin
dependence will help to solve this $\delta_{LT}$ puzzle.

The generalized longitudinal-transverse
polarizability can be extracted from a measurement of the spin structure
functions $g_1$ and $g_2$:
\begin{eqnarray}
\delta_{LT}(Q^2)&=&
\frac{16 \alpha M^2}{Q^6}\int^{x_0}_0 x^2 \Bigl [g_1(x,Q^2)+g_2(x,Q^2)
\Bigr ] dx.
\end{eqnarray}
Measurement of $g_2$ also allows for a test of the
Burkhardt-Cottingham (BC) sum rule~\cite{BC},
a ``super-convergence
relation'' that is valid for any value of $Q^2$:
\begin{equation}
\label{eq:bc}
\int_0^{1}g_2(x,Q^2)dx=0,
\end{equation}

E07-001~\cite{DLT} has been conditionally approved to run in JLab Hall A.
We will perform an inclusive  measurement at forward angle of the proton
spin-dependent cross sections in order to
determine the
$g_2^p$ structure function in the resonance region for $0.02<Q^2<0.4$ GeV$^2$.
This measurement will allow an extraction of
$\delta_{LT}$, and
a test of the BC sum rule.

\begin{figure}
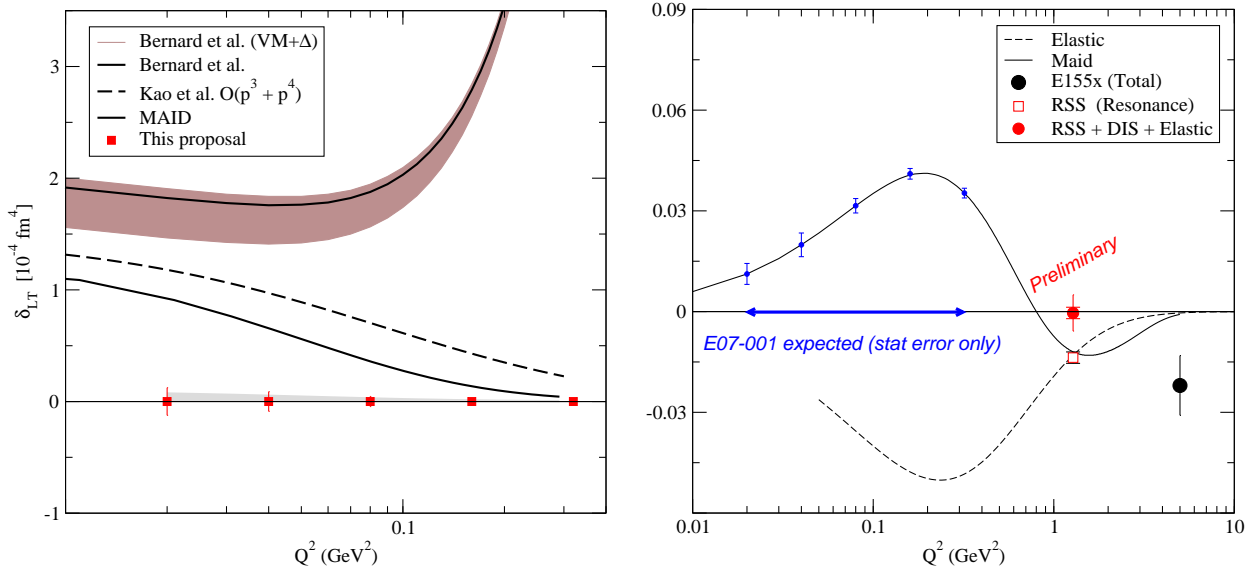

\includegraphics[width=0.485\textwidth]{figs/dlt_bernard_log.eps}
\hspace{0.2cm}
\includegraphics[width=0.485\textwidth]{figs/gam2_log_07001b.eps}
\caption{\label{PRED} {\bf Left: } Projected results for $\delta_{LT}$. Statistical errors are shown on the symbols.  Systematic is represented by the grey band
on the axis.
$\chi$PT predictions from Bernard {\it et al.}~\cite{ber},
and Kao {\it et al.}~\cite{van02}.
{\bf Right:} Projected statistical error bars for $\Gamma_2^p(Q^2)$.
}
\end{figure}

For this experiment we will install the JLab-UVA polarized
ammonia target in Hall A for the first time. This installation will
involve significant changes to the existing Hall A beamline in order
to properly transport the electron beam in the presence of the 5T
magnetic field of the transversely polarized target.
Projected results for $\delta_{LT}$ and the BC sum rule are shown in
Fig~\ref{PRED}.

\section{RSS: The Resonant Spin Structure Experiment}
Experiment E01-006 was conducted in Hall C of the
Thomas Jefferson National Accelerator Facility. 
We  measured
parallel and perpedicular proton and deuteron asymmetries  in the scattering of 100 nA polarized
5.755 GeV electrons.
Scattered electrons were detected
at an
angle of 13.15$^\circ$ using the Hall C High Momentum Spectrometer.
Full details of the experiment
can be found in Refs.~\cite{PRL} and~\cite{Jones:2006kf}.

The twist-3 matrix element d$_2$ is related to the higher twist contribution to
$g_2$ via:

\begin{eqnarray}
\label{D2EQ}
d_2(Q^2) = 3\int_0^1 x^2\left[g_2(x,Q^2) - g_2^\mathtt{WW}(x,Q^2)\right] dx
\end{eqnarray}
where $g_2^\mathtt{WW}(x,Q^2$) is the leading twist part of $g_2$ and is given by 
the Wandzura-Wilczek~\cite{WW} relation:
\begin{eqnarray}
\label{G2WWEQ}
g_2^\mathtt{WW}(x,Q^2) = -g_1(x,Q^2) + \int_x^1 \frac{g_1(y,Q^2)}{y} dy
\end{eqnarray}

\begin{figure}
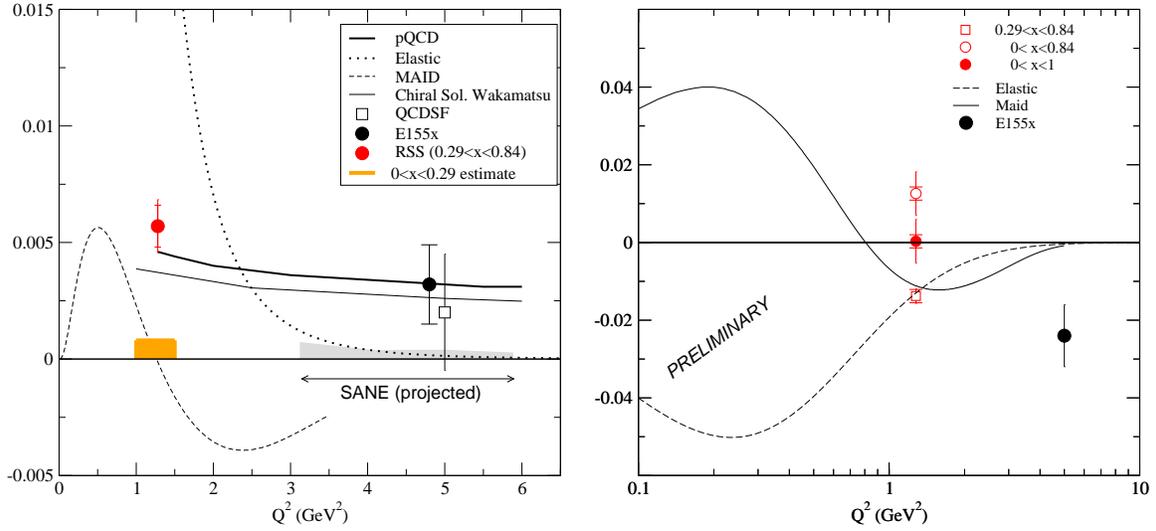
\begin{center}\includegraphics[width=0.45\textwidth]{figs/d2.eps}
\hspace{0.1cm}
\includegraphics[width=0.45\textwidth]{figs/gam2p_log_2.eps}
\end{center}
\caption{RSS results. 
{\bf Left: } d$_2^p(Q^2=1.279)$~\cite{PRL}.
{\bf Right:} $\Gamma_2^p$(Q$^2$).}
\label{GAM2}\end{figure}

Using ~\ref{G2WWEQ}, we can re--express Eq.~\ref{D2EQ} as a higher moment of a simple linear combination
of the g$_1$ and g$_2$ structure functions:
\begin{eqnarray}
d_2(Q^2) = \int_0^1 x^2\left[2 g_1(x,Q^2) + 3 g_2(x,Q^2)\right] dx
\end{eqnarray}

d$_2(Q^2)$ measures the deviation of
the g$_2$ structure function from leading twist behaviour,
 and as such, it is
sensitive to quark-gluon interactions beyond the simple quark parton model.
Fig.~\ref{GAM2} shows the measured value~\cite{PRL} of $d_2(Q^2=1.279)$ in the left panel.
Our results are clearly non-zero by several standard deviations.

The right panel displays $\Gamma_2=\int g_2(x,Q^2) dx$.  
The RSS resonance region
data covers the range  $1090 < W < 1910$ MeV, ($0.3161 < x < 0.823$).
To estimate the unmeasured contribution to $\Gamma_2$ as $x\to0$ we assume
that the Wandzura-Wilczek~\cite{WW} relation holds
and that
\begin{eqnarray}
\label{WW1}
\int_0^{x_0} g_2(x,Q^2) dx &\simeq& \int_0^{x_0} g_2^\mathtt{WW}(x,Q^2) dx \\
\label{WW2}
                       &=& x_0 \int_{x_0}^1 \frac{g_1(x,Q^2)}{x} dx
\end{eqnarray}
where Eq.~\ref{WW2} follows from integration by
parts.
The total integral exhibits a striking cancellation of the inelastic (resonance+DIS) and elastic contributions,
leading to satisfaction of the Burkhardt-Cottingham sum rule within
uncertainties.

\begin{figure}
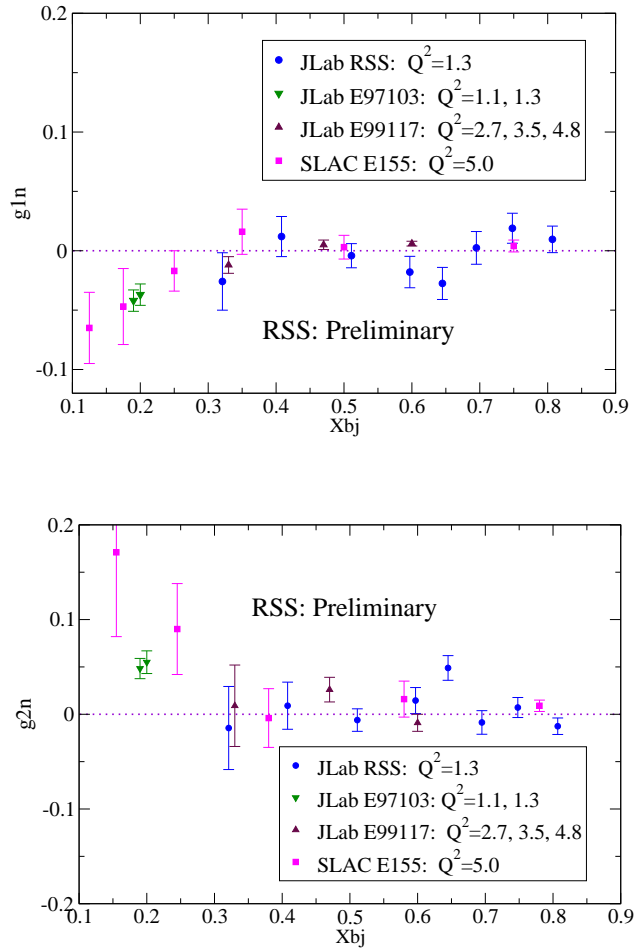
\begin{center}\includegraphics[width=0.5\textwidth]{figs/g1n_compare.eps}
\vspace{1cm}

\includegraphics[width=0.5\textwidth]{figs/g2n_compare.eps}
\end{center}
\caption{Preliminary RSS Neutron $g_1$ and $g_2$ structure functions compared to world data.}
\label{SHIGE}\end{figure}

The combination of high precicion proton and deuteron data also allows the extraction of
the neutron structure fuctions as shown in Fig.~\ref{SHIGE}.


\section{Spin Asymmetries of the Nucleon Experiment  }
\begin{figure}\begin{center}
\includegraphics[width=0.8\textwidth]{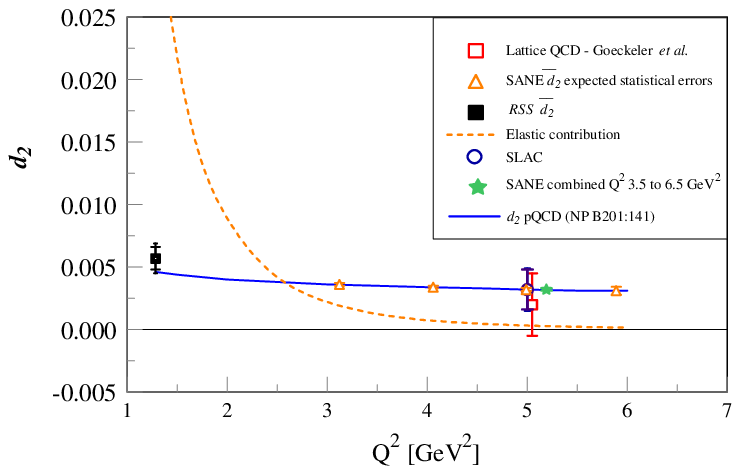}
\end{center}
\caption{SANE: Projected errors for proton $d_2$. }
\label{SANEFIG2}\end{figure}

\begin{figure}\begin{center}
\includegraphics[width=0.45\textwidth]{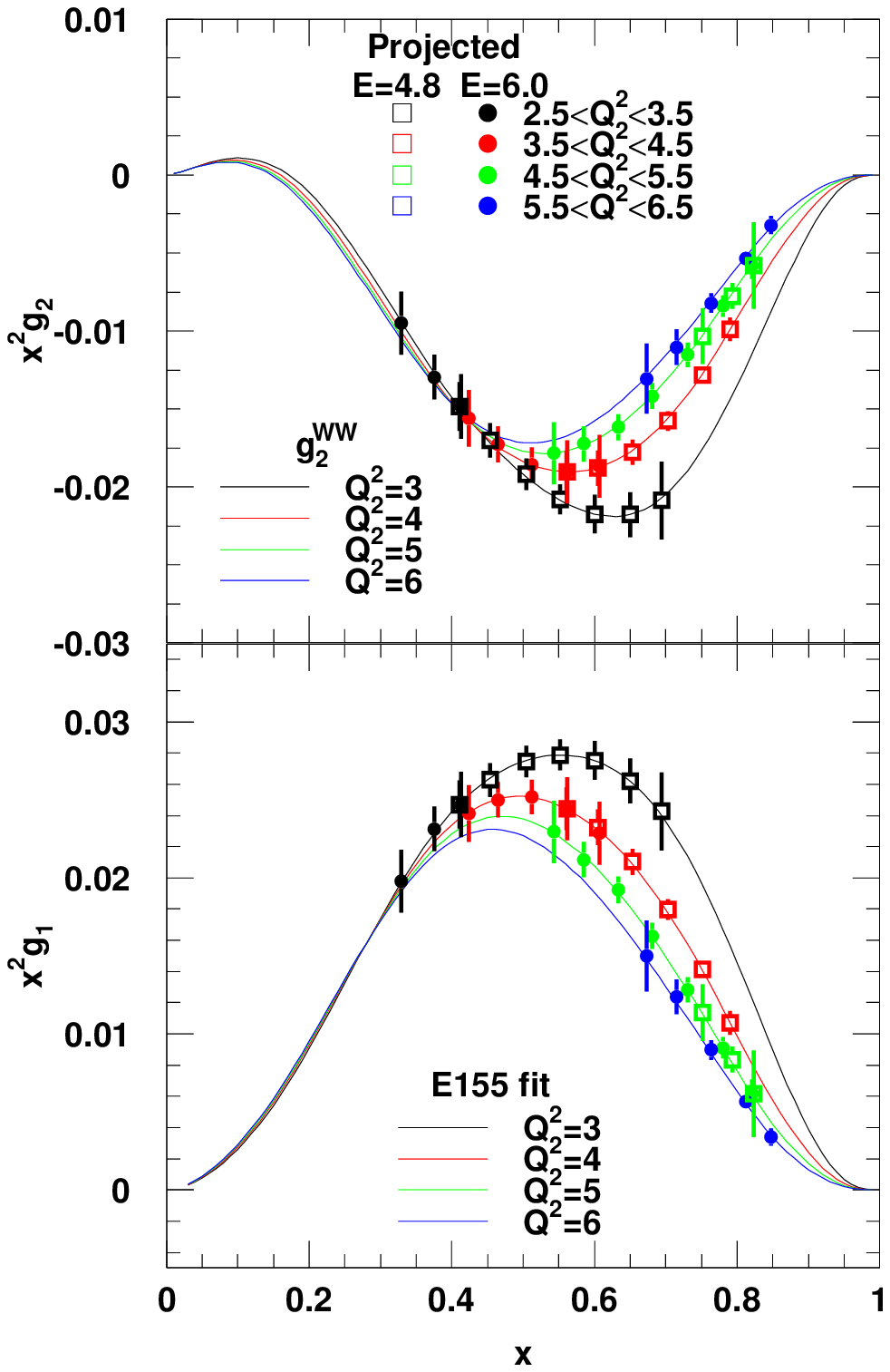}
\end{center}
\caption{SANE: Projected errors for proton $g_1$, $g_2$ }
\label{SANEFIG1}\end{figure}

JLab E07-003 is scheduled to run in Hall C for 70 days in 2008.
The SANE experiment will perform an inclusive double polarization measurement
in order to extract 
the proton spin structure functions $g_1$ and $g_2$  
for 
$0.3 \le x \le 0.8$ and $2.5$ GeV$^2 \le Q^2 \le$ 6.5
GeV$^2$, in a region where the nucleon spin structure remains largely unknown.
Higher twists will be investigated via the moments
of $g_1$ and $g_2$.  Expected precision is shown in Fig.~\ref{SANEFIG2} and~\ref{SANEFIG1}.

\section{Summary}
We have presented here a small sample of the high quality Jefferson Lab spin program
data.  For a more general overview, 
we refer the reader to Ref.~\cite{Chen:2005tda}.  
We look forward to many future discoveries and challenges as the JLab 12 GeV upgrade~\cite{12GEV}
commences.


\end{document}